\documentclass[trackchanges]{aastex7}

\usepackage{float}
\usepackage{graphicx}
\graphicspath{{E:/apj/apj/figures/}}
\usepackage{natbib}
\usepackage{multirow}
\usepackage{epstopdf}
\shorttitle{EUV wave and QPPs during an M-class flare}
\shortauthors{Li et al.}

\begin{document}
	
	\title{Extreme Ultraviolet Wave and Quasi-periodic Pulsations during an eruptive M-class Flare}
	
	\correspondingauthor{Qingmin Zhang}
	\email{zhangqm@pmo.ac.cn}
	
	\author[0009-0000-5839-1233]{Shuyue Li}
	\affiliation{Purple Mountain Observatory, Chinese Academy of Sciences, Nanjing 210023, People's Republic of China}
	\affiliation{School of Science, Nanjing University of Posts and Telecommunications, Nanjing 210023, People's Republic of China}
	\email{syli@pmo.ac.cn}
	
	\author[0000-0003-4078-2265]{Qingmin Zhang}
	\affiliation{Purple Mountain Observatory, Chinese Academy of Sciences, Nanjing 210023, People's Republic of China}
	\email{zhangqm@pmo.ac.cn}
	
	\author[0000-0002-5898-2284]{Haisheng Ji}
	\affiliation{Purple Mountain Observatory, Chinese Academy of Sciences, Nanjing 210023, People's Republic of China}
	\email{jihs@pmo.ac.cn}
	
	\author{Shengli Liu}
	\affiliation{School of Science and New Energy Technology Engineering Laboratory of Jiangsu Province, 
	Nanjing University of Posts and Telecommunications, Nanjing 210023, People's Republic of China}
	\email{liusl@njupt.edu.cn}
	
	\author[0000-0003-0779-3712]{Fanpeng Shi}
	\affiliation{School of Astronomy and Space Science, Nanjing University, Nanjing 210093, People's Republic of China}
	\email{f.shi@nju.edu.cn}
	
	\author[0000-0002-4538-9350]{Dong Li}
	\affiliation{Purple Mountain Observatory, Chinese Academy of Sciences, Nanjing 210023, People's Republic of China}
	\email{lidong@pmo.ac.cn}

\begin{abstract}
In this paper, we report multiwavelength and multipoint observations of the prominence eruption originating from active region 11163, 
which generated an M3.5 class flare and a coronal mass ejection (CME) on 2011 February 24. The prominence lifts off and propagates nonradially in the southeast direction.
Using the revised cone model, we carry out three-dimensional reconstructions of the icecream-like prominence.
It is found that the latitudinal inclination angle decreases from $\sim$60$\degr$ to $\sim$37$\degr$, indicating that the prominence tends to propagate more radially. 
The longitudinal inclination angle almost keeps constant (-6$\degr$).
The highly inclined prominence eruption and the related CME drive an extreme ultraviolet (EUV) wave,
which propagates southward at speeds of $\sim$381.60 km s$^{-1}$ and $\sim$398.59 km s$^{-1}$ observed in 193 {\AA} and 304 {\AA}, respectively.
The M3.5 class flare presents quasi-periodic pulsations (QPPs) in soft X-ray, hard X-ray, EUV, and radio wavelengths with periods of 80$-$120 s.
Cotemporary with the flare QPPs, a thin current sheet and multiple plasmoids are observed following the eruptive prominence.
Combining with the appearance of drifting pulsation structure,
the QPPs are most probably generated by quasi-periodic magnetic reconnection and particle accelerations as a result of plasmoids in the current sheet.
\end{abstract}
	

\section{Introduction} \label{intro}
Solar flares are impulsive increases of electromagnetic emissions in multiple wavelengths \citep{car59,fle11,jan15}.
The accumulated magnetic free energy in active regions (ARs) is drastically released via magnetic reconnection within a few to tens of minutes 
and converted into radiations, thermal and kinetic energies of plasmas, as well as nonthermal energy of accelerated particles \citep{pri02,dra25}. 
For eruptive flares associated with coronal mass ejections \citep[CMEs;][]{for06,chen11}, 
a significant part of magnetic energy is converted into the kinetic energy of CMEs \citep{ems12,zqm23}.
Coronal extreme ultraviolet (EUV) waves could be generated by flares \citep{shen12a}, CMEs \citep{cw11,pan25}, and coronal jets \citep{shen18,hou23}.

Quasi-periodic pulsations (QPPs) are quasi-periodic variations in radiations \citep[see reviews][and references therein]{naka09,zim21}.
They are quite commonplace not only in solar flares but also in stellar flares \citep{ing15,ing16,hay20,ing24,jos25}.
\citet{par69} reported the first QPP event with a period of $\sim$16 s detected in hard X-ray (HXR) on 1968 August 8.
The peaks in HXR are accordant with those in microwave (15.4 GHz). 
\citet{kan83} concluded that the observed variations in HXR and microwave emissions are due to time variations in the electron acceleration/injection spectrum.
QPPs are observed over a wide range of wavelengths, from radio, Ly$\alpha$, white light (WL), extreme-ultraviolet (EUV) to X-rays, 
and even $\gamma$-rays \citep{ing08,naka10,lid15,cla21,fre24,lid24,lid25}. 
The periods range from subseconds \citep{knu20} to several minutes \citep{ning22,col23,pur25}, 
most of which are between seconds to tens of seconds \citep{zqm16,chen19,kou22,fre24}.
\citet{tan16} reported very long period (8$-$30 minutes) QPPs, which last for 1$-$2 hr in the preflare phase of flares.
Occasionally, the period of QPP increases with time as the flare evolves \citep{rez11,den17,hay16,hay19,col24}.
Multiple periods of QPPs are frequently detected in a single flare \citep{ing09,tan10,zim10,zqm22b,col23,shi25}.
\citet{ing09} studied a multi-periodic event with three distinct periods (28, 18, and 12 s). Considering the period ratios of this event, 
they suggested that the possible cause of this phenomenon is a kink mode that periodically stimulates the magnetic reconnection.
\citet{song25} discovered harmonic QPPs ($\sim$11 s and $\sim$20 s) in 3600 {\AA} continuum of an X2.8 white-light (WL) flare on 2023 December 14.
The spatial sources of QPPs are at the flare ribbons or kernels in most cases \citep{zqm16,cla21,lid25,song25}.
\citet{shi25} investigated a looptop HXR source, which exhibits an oscillation in height that is anticorrelated with the HXR intensity on 2022 August 26.
A period of $\sim$2 minutes with a large amplitude and a relatively weak period of $\sim$1 minute are found in the QPPs.
Besides, the non-thermal electron spectral index and flux as well as the EUV intensity of flare/ribbons show similar oscillations.

Although observations and numerical simulations of QPPs are extensive, 
the fundamental physical mechanisms underlying QPPs remain a topic of debate \citep{kup20,zim21}.
The dominant mechanism is periodic or intermittent magnetic reconnection and particle acceleration \citep{ing08,lid15}.
The self-oscillatory process includes spontaneous repetitive reconnection, such as oscillatory reconnection, 
where inflow and outflow regions change regularly \citep{mur09,mc09,hjc19,hjc25}.
Another source of periodicity is intermittent formations of plasmoids as a result of tearing and plasmoid instabilities 
in a thin current sheet during magnetic reconnection \citep{shi01,asai04,lin05,bar08,tak12,ni15,ni17,cx18,kou22,lu22,yan22,hou24,kum25}.
The particles are efficiently accelerated as the plasmoids contract \citep{dra06}.
Besides, the reconnection rate could be modulated by external magnetohydrodynamic (MHD) waves, 
such as fast-mode waves \citep{naka06} and slow-mode waves \citep{sych09}.
\citet{lid24} studied the QPPs during the impulsive phase of an X6.4 flare on 2024 February 22.
The long period ($\sim$200 s) is probably modulated by the slow mode wave from the nearby sunspot, 
while the shorter period ($\sim$95 s) is regarded as the second harmonic mode.
\citet{zqm16} explored the QPPs during the impulsive phase of a C3.1 circular-ribbon flare on 2015 October 16.
According to the flare size and observed periods (32$-$42 s) in UV and SXR derivative,
it is concluded that the QPPs are most probably caused by the fast wave modulation of magnetic reconnection, particle acceleration and injection into the chrompshere.
\citet{tian18} discovered repetitive chromospheric condensation as a result of episodic heating by nonthermal electrons 
on a timescale of 1$-$3 minutes during the M1.6 flare on 2015 March 12.
\citet{hay16} investigated the QPPs during the impulsive and decay phases of the X1.0 flare on 2013 October 28.
The QPPs during the impulsive phase are interpreted by episodic particle acceleration and plasma heating, 
while the thermal decay phase QPPs are interpreted by compressive MHD waves of fast sausage mode or vertical kink mode in the post-flare loops. 

Quasi-periodic fast-propagating (QFP) magnetosonic wave trains were first discovered in a C3.2 class flare/CME event on 2010 August 1 \citep{liu11}.
The speeds and main period of QFP are $\sim$2200 km s$^{-1}$ and $\sim$181 s for that event.
Many studies suggest that QPPs and QFPs are different manifestations of the same underlying process, i.e., periodic energy releases via magnetic reconnection \citep{shen22}.
\citet{zhou21} studied the CME-driven and flare-ignited fast magnetosonic waves on 2013 April 23. 
The periods (125 s and 85 s) of narrow quasi-periodic EUV wave are comparable to the period (113 s) of associated flare QPP.
\citet{zhou22} investigated a flare-driven quasi-periodic EUV wave train on 2015 December 22. 
The periods (102$\pm$1 s) of flare QPP in SXR are comparable to the periods (84$-$110 s) of wave train.

\citet{kum12} analyzed the M3.5 flare associated with a CME in AR NOAA 11163 (N19E77) near the eastern limb on 2011 February 24.
The eruptive flare results from a typical nonradial eruption of a twisted magnetic flux rope due to helical kink instability \citep{ji03,tor04}.
Two bright structures (blob ``A" and blob ``B") are identified in the flux rope 
and the speed profile of blob ``A" shows quasi-periodic oscillations with a period of $\sim$2 minutes.
In this paper, we revisit this event, focusing on the EUV wave as a result of the nonradial eruption and QPPs of the flare.
The paper is organized as follows. We describe the instruments and data analysis in Section~\ref{instr}.
In Section~\ref{res}, the first part shows the prominence eruption and associated EUV wave,
and the second part shows the flare QPPs and plasmoids in the current sheet beneath the flux rope.
Discussions and a brief summary are given in Section~\ref{dis} and Section~\ref{sum}, respectively.

\begin{deluxetable*}{ccccccc}
		\digitalasset
		\tablewidth{\textwidth}
		\tablecaption{Properties of the instruments used in this study.
			\label{tab1}}
		\tablecolumns{4}
		\tablenum{1}
		\tablehead{
			\colhead{Instrument} &
			\colhead{Wavelength} &
			\colhead{Cadence} &
			\colhead{Pixel Size}  \\
			\colhead{ } &
			\colhead{ } &
			\colhead{(s)} &
			\colhead{(arcsec)} &
		}
		\startdata
		 SDO/AIA & 131, 171, 193, 211, 304 {\AA} & 12 & 0.6   \\
		 LASCO-C2 & WL & 720 & 11.4 \\
         STB/EUVI & 304 {\AA} & 600 & 1.6  \\
		 GOES/XRS & 0.5$-$4 {\AA}, 1$-$8 {\AA} & 1 & -  \\
		 RHESSI & 12$-$300 keV & 4 & - \\
		 BLEN7M & 175$-$870 MHz & 0.25 & -
		\enddata
	\end{deluxetable*}

\section{Instruments and data analysis} \label{instr}
\begin{figure} 
   \includegraphics[width=0.45\textwidth,clip=]{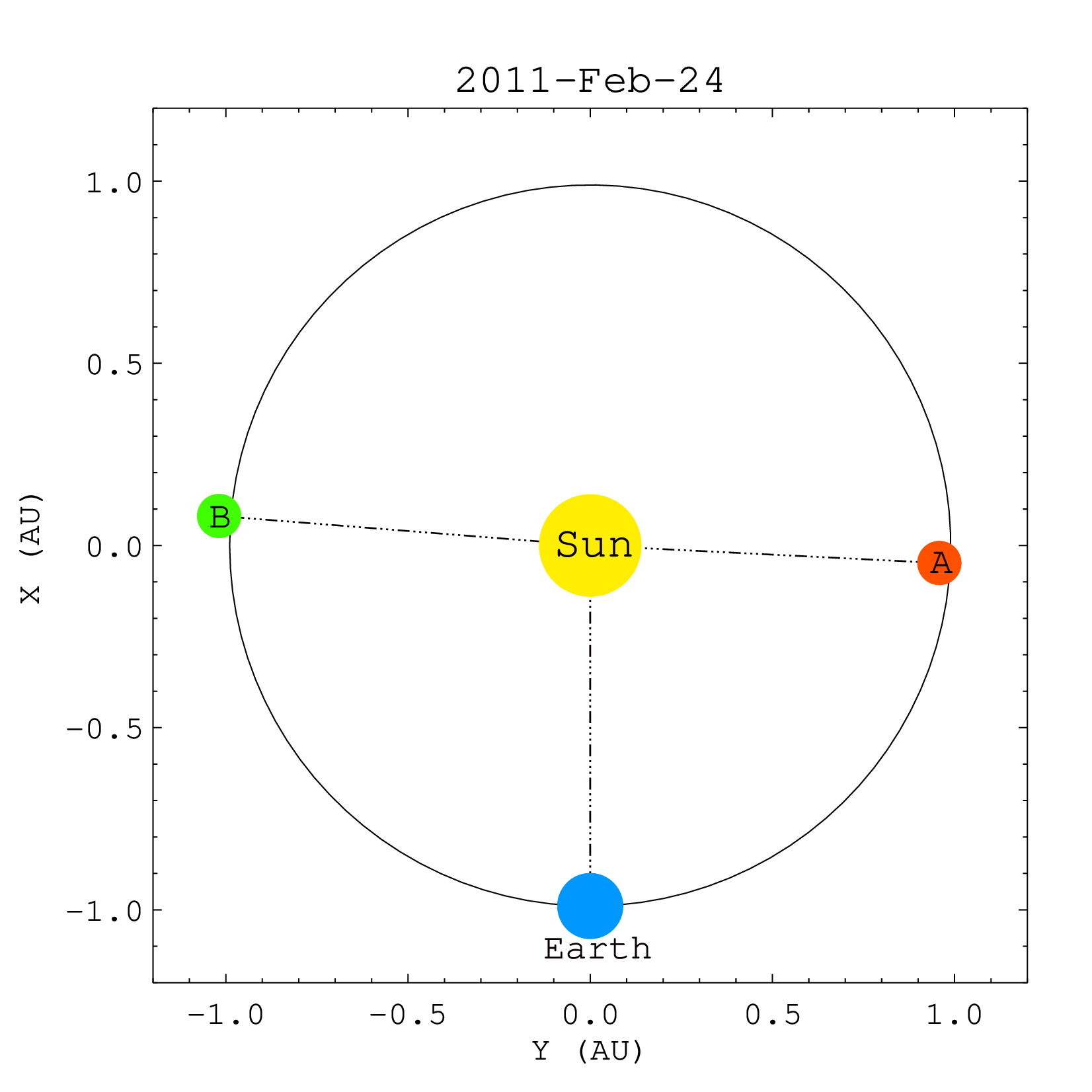}
   \centering
    \caption{The locations of STA (red circle), STB (green circle), and Earth (blue circle) on 2011 February 24. 
    		The separation angles of STA and STB with the Sun-Earth connection are 87$\fdg$1 and 94$\fdg$5, respectively.}
    \label{fig1}
    \end{figure}

\begin{figure} 
		\includegraphics[width=0.85\textwidth]{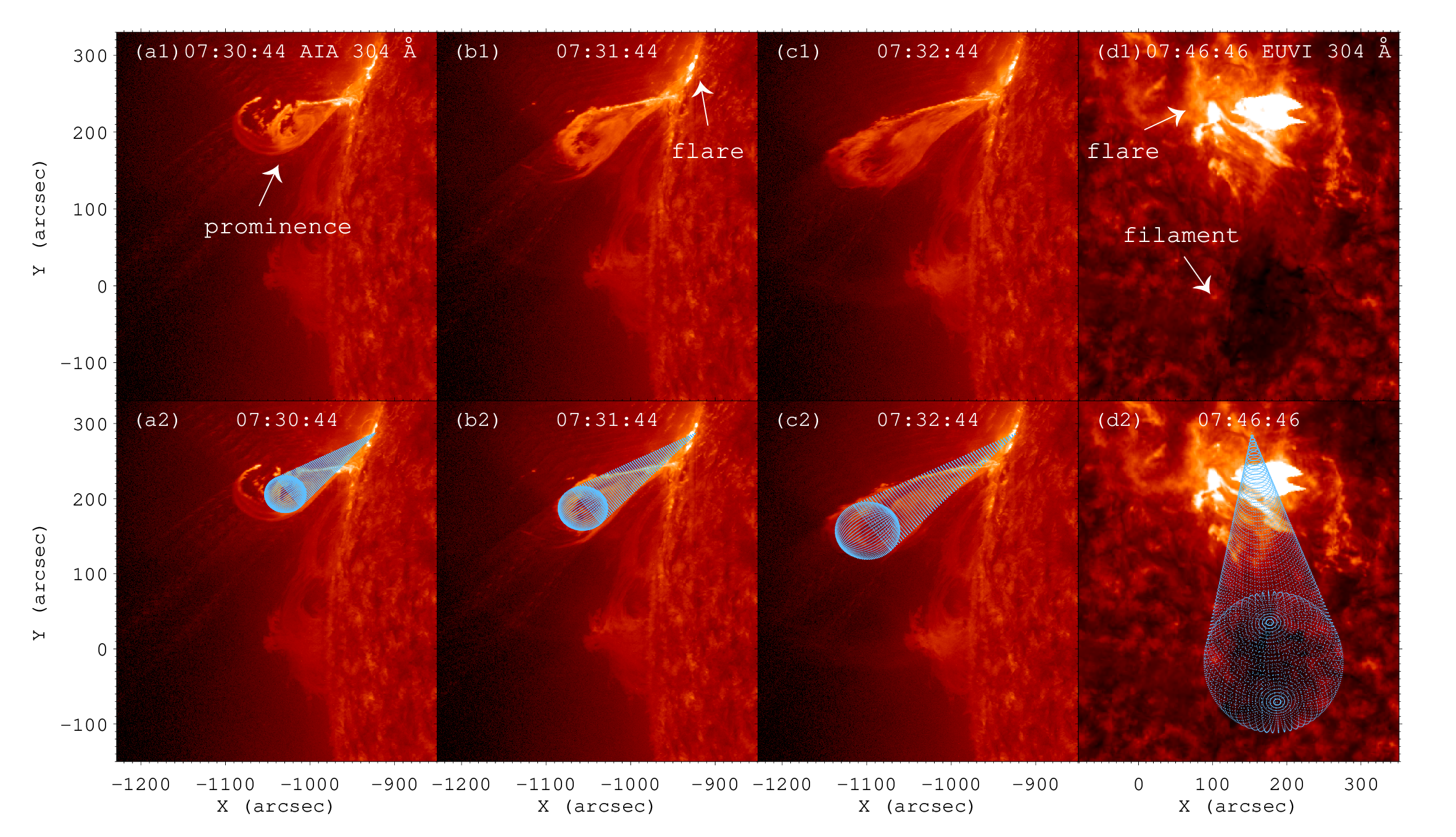}
		\centering
		\caption{(a1)-(c1) snapshots of the AIA 304 {\AA} images during the nonradial eruption. 
		(a2)-(c2) same images as the top panels, with the projections of the reconstructed cones being superposed (blue dots).
		(d1) the eruptive prominence observed by STB/EUVI at 07:46:46 UT.
		(d2) the same image as (d1), with the projection of the reconstructed cone being superposed (blue dots).
		An animation showing the prominence eruption observed by AIA 304 {\AA} is available.
		It covers a duration of 80 minutes from 07:20 UT to 08:40 UT on 2011 Feburary 24. The entire animation runs for $\sim$5 s.
		(An animation of this figure is available in the online article.)}
		\label{fig2}
	\end{figure}

The eruptive prominence and flare were observed by several telescopes from multiple viewpoints, 
including the Atmospheric Imaging Assembly \citep[AIA;][]{lem12} on board the Solar Dynamics Observatory \citep[SDO;][]{pes12}
and the Extreme UltraViolet Imager (EUVI) of the Sun Earth Connection Coronal and Heliospheric Investigation \citep[SECCHI;][]{how08}
on board the Solar TErrestrial RElations Observatory \citep[STEREO;][]{kai08} behind (hereafter STB).
Figure~\ref{fig1} shows the locations of the ahead STEREO (hereafter STA), STB, and Earth with red, green, and blue circles, respectively.
The separation angles of STA and STB with the Sun-Earth connection were 87$\fdg$1 and 94$\fdg$5.
Therefore, the eruptive prominence was above the eastern limb and on the disk in the field of views (FOVs) of SDO/AIA and STB/EUVI, respectively.
It was on the farside of the Sun as viewed from STA.
The upper panels of Figure~\ref{fig2} show the bright prominence observed by SDO/AIA during 07:30$-$07:32 UT (panels (a1)-(c1))
and the dark filament observed by STB/EUVI at 07:46:46 UT (panel (d1)).
The prominence starts to rise at $\sim$07:22 UT and propagates nonradially in the southeast direction.
The top of prominence escapes the FOV of AIA after $\sim$07:34 UT, while the trailing materials are still visible until $\sim$08:30 UT 
(see also the online animation \textit{anim1.mp4}).

\begin{figure} 
\includegraphics[width=0.55\textwidth,clip=]{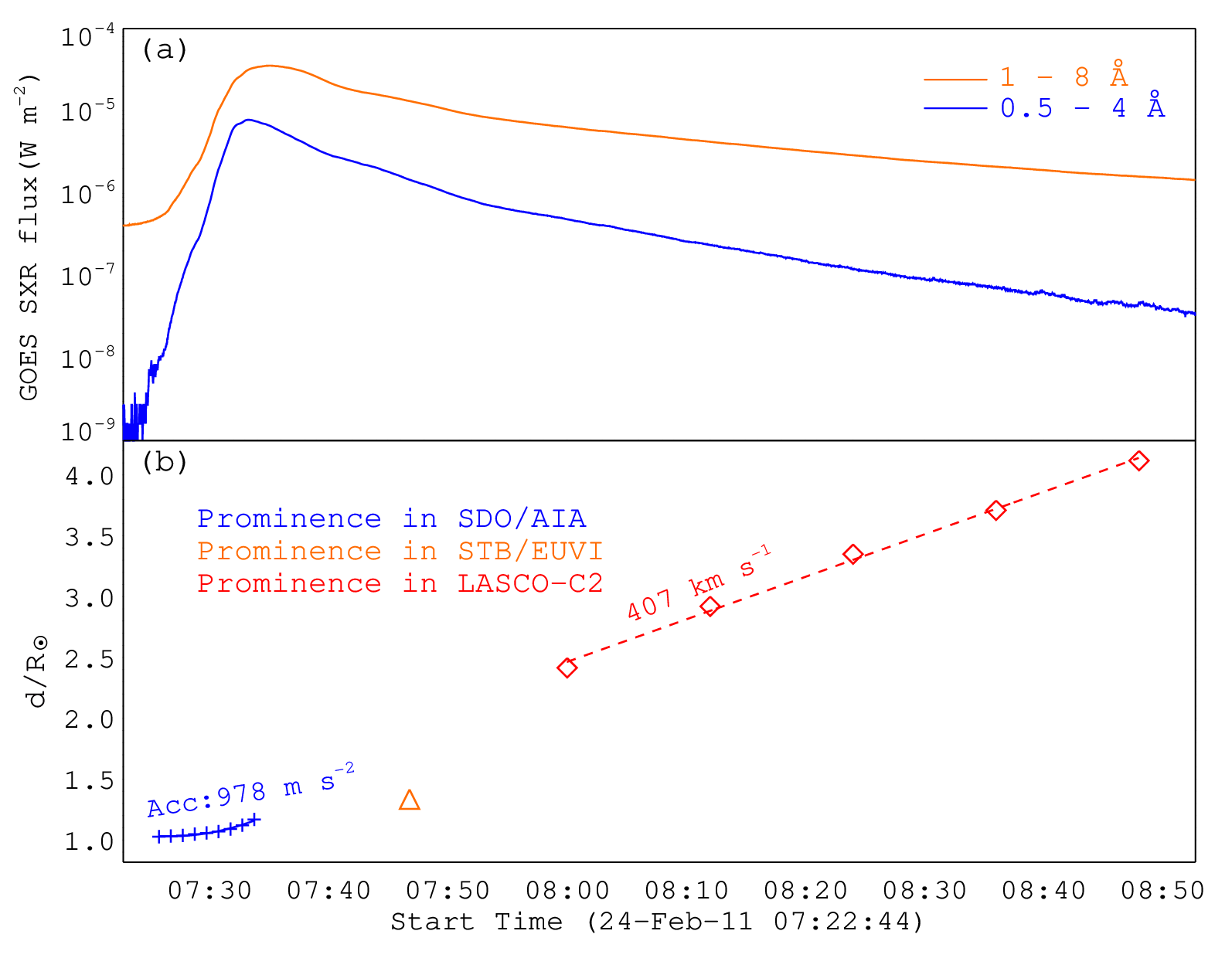}
\centering
\caption{(a) SXR light curves of the flare in 1$-$8 {\AA} (orange line) and 0.5$-$4 {\AA} (blue line). 
              (b) time evolution of the heliocentric distances of the prominence leading front 
              in the FOV of AIA (blue pluses), STB/EUVI (orange triangle), and in LASCO (red diamonds).}
	     \label{fig3}
\end{figure}

The top panel of Figure~\ref{fig3} shows SXR light curves of the M3.5 flare in 1$-$8 {\AA} (orange line) and 0.5$-$4 {\AA} (blue line) 
recorded by the X-Ray Sensor \citep[XRS;][]{han96} on board the Geostationary Operational Environmental Satellite \citep[GOES;][]{gar94} spacecraft.
The flare started at 07:23 UT, peaked at 07:35 UT, and ended at 07:42 UT \citep{kum12}. The start times of flare and prominence eruption are very close.
HXR light curves of the flare at different energy bands were observed by the Reuven Ramaty High-Energy Solar Spectroscopic Imager \citep[RHESSI;][]{lin02}. 

\begin{figure} 
		\includegraphics[width=0.85\textwidth]{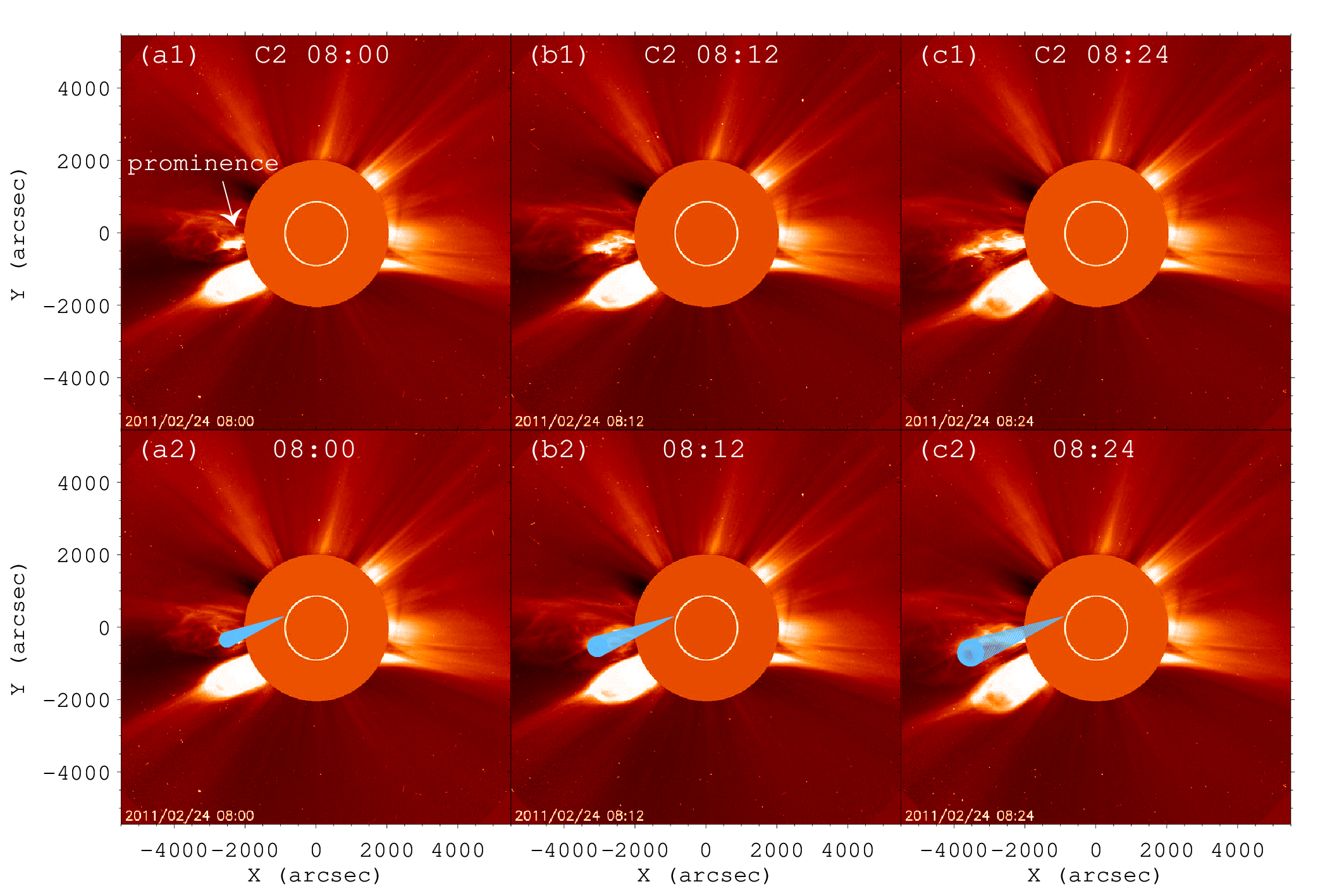}
		\centering
		\caption{(a1)-(c1) WL images of the CME observed by LASCO-C2 during 08:00$-$08:24 UT.
		(a2)-(c2) the same images as the top panels, with the projections of reconstructed cones being superposed with blue dots.}
		\label{fig4}
\end{figure}

\begin{figure} 
		\includegraphics[width=0.5\textwidth]{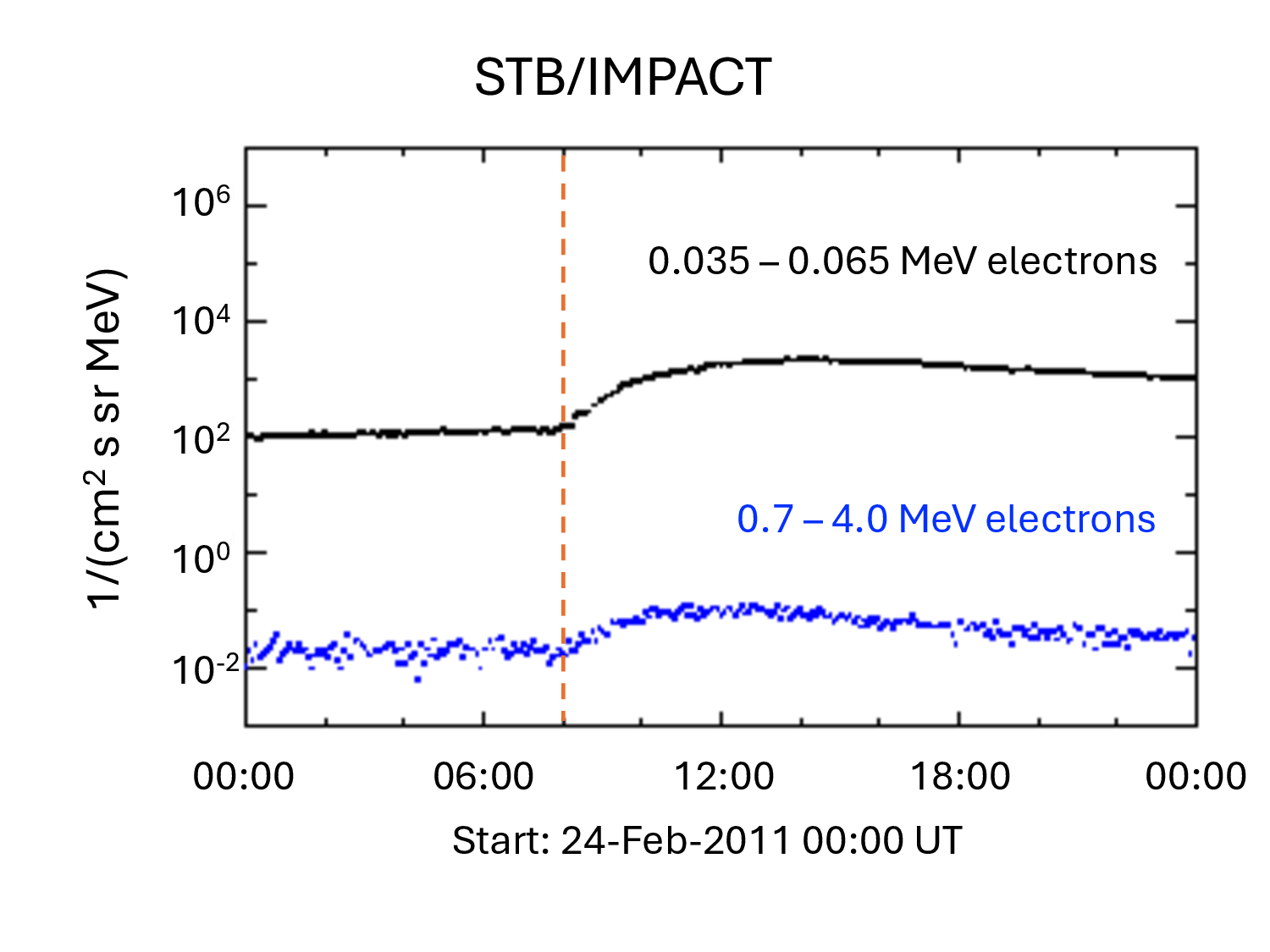}
		\centering
		\caption{Electron fluxes recored by STB/IMPACT.
		The vertical orange line denotes 08:00 UT.}
		\label{fig5}
	\end{figure}

The related CME driven by the prominence eruption was observed by the C2 coronagraph of the Large Angle Spectroscopic Coronagraph \citep[LASCO;][]{bru95} 
on board the Solar and Heliospheric Observatory (SOHO) mission. The top panels of Figure~\ref{fig4} show the CME observed by LASCO-C2 during 08:00$-$08:24 UT.
In panel (a1), the white arrow points to the bright core (prominence) of the CME with a typical three-part structure \citep{lsy25}.
Although the CME leading edge first appears at 07:48 UT in the FOV of LASCO-C2, the following prominence shows up at $\sim$08:00 UT.
The position angle, angular width, and linear speed of the CME in the plane-of-the-sky (POS) are $\sim$70$\degr$, $\sim$158$\degr$, and $\sim$1186 km s$^{-1}$
in the CDAW website\footnote{https://cdaw.gsfc.nasa.gov/CME\_list/UNIVERSAL\_ver2/2011\_02/univ2011\_02.html}.
As displayed in their Fig. 5 \citep{kum12}, a type II radio burst was observed during 07:37$-$08:05 UT, indicating a coronal piston-driven shock by the fast CME. 
The shock was potentially capable of accelerating electrons to high energies, generating the solar energetic particle \citep[SEP;][]{rea13,des16,rod25} event.
Figure~\ref{fig5} shows the time evolution of electron fluxes with energies of 0.036$-$0.065 MeV (black line) and 0.7$-$4.0 MeV (blue line) 
detected by the In-situ Measurements of Particles and CME Transient \citep[IMPACT;][]{luh08} on board STB on 2011 February 24.
The vertical orange line denotes 08:00 UT, when the electron fluxes start to increase gradually.

Radio dynamic spectra of the flare in the frequency range of 175$-$870 MHz, 
featuring a drifting pulsation structure \citep[DPS;][]{kli00,kar04,nis15,hua16,kar18,lu22},
were observed by the radio telescope at e-Callisto\footnote{https://www.e-callisto.org} \citep{benz09} BLEN7M.
The light curves in SXR, HXR, EUV, and radio wavelengths are first smoothed with a window of 160 s to derive the slowly-varying components. 
After subtracting the slowly-varying components, the fast-varying components (detrended light curves) are obtained 
to investigate QPPs of the flare using the Morlet wavelet transform \citep{dau90,lid15}, which will be presented in Section~\ref{qpp}.
The properties of instruments and their data used in this study are summarized in Table~\ref{tab1}.

\section{Results} \label{res}

\subsection{Prominence eruption and EUV wave} \label{pe}
To investigate the 3D morphology and trajectory of nonradial solar eruptions \citep{guo23,chen24}, \citet{zqm21} put forward a revised cone model, 
in which the tip of the cone is located at the source region of the eruption rather than the Sun center. The model is characterized by four parameters:
the length ($r$) and angular width ($\omega$) of the cone, latitudinal ($\theta_1$) and longitudinal ($\phi_1$) deflection angles of the cone relative to the local vertical.
$\theta_1$ is positive (negative) when the eruption is inclined southward (northward), respectively.
$\phi_1$ is positive (negative) when the eruption is inclined westward (eastward), respectively.
For an exactly radial eruption, $\theta_1=0$ and $\phi_1=0$. 
For a cone with a spherical leading front, which resembles an ice cream, the height of the front is \citep{zqm22a}:
\begin{equation} \label{eqn-1}
l=r[\tan{\frac{\omega}{2}}+(\cos{\frac{\omega}{2}})^{-1}].
\end{equation}

Using multiwavelength and multipoint observations, the model has been successfully applied to the 3D reconstructions and tracking of leading edges of 
CMEs \citep{zqm21,zqm22a} and prominences \citep{zqm24}.
Deflection, expansion, and acceleration in their early phases are nicely revealed, which are important to the space weather forecast.
As shown in the top panels of Figure~\ref{fig2}, the prominence propagates nonradially in the southeast direction.
We carry out 3D reconstructions for the prominence using the revised cone model and observations from SDO/AIA, STB/EUVI, and LASCO-C2.
In Figure~\ref{fig2}, the bottom panels show the same EUV 304 {\AA} images as the top panels.
The projections of the reconstructed cones are superposed with blue dots.
It is seen that the tips of the cones are located in AR 11163 and the cones fit well with the prominence (filament) body.
In Figure~\ref{fig4}, the bottom panels show the same WL images of CME as the top panels. Likewise, the projections of the cones are superposed with blue dots.
It should be emphasized that we did not perform reconstructions for the CME leading edge.

\begin{figure} 
		\includegraphics[width=0.90\textwidth]{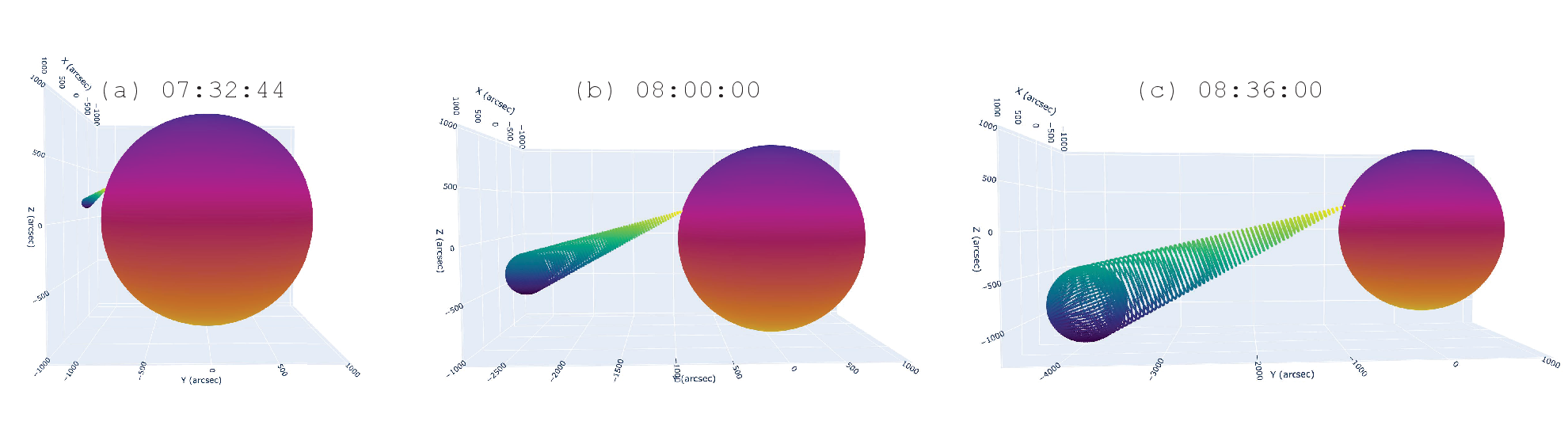}
		\centering
		\caption{3D visualization of the Sun and reconstructed cones at three moments.
		The YZ plane defines the plane of the sky.}
		\label{fig6}
	\end{figure}

A total of 15 moments are utilized to carry out prominence reconstructions from 07:25 UT observed by AIA to 08:48 UT observed by LASCO-C2.
Figure~\ref{fig6} demonstrates 3D visualization of the Sun and cones at 07:32 UT (a), 08:00 UT (b), and 08:36 UT (c).
In Table~\ref{tab2}, the parameters of these cones are listed.
The value of $r$ increases from $\sim$57$\arcsec$ at 07:25 UT to $\sim$290$\arcsec$ at 07:33 UT before escaping the FOV of SDO/AIA.
It increases drastically from $\sim$1700$\arcsec$ to $\sim$3630$\arcsec$ in the FOV of LASCO-C2.
The corresponding value of $l$ (Equation~\ref{eqn-1}) increases from $\sim$64$\arcsec$ to $\sim$4150$\arcsec$.
The latitudinal inclination angle $\theta_1$ decreases monotonously from $\sim$60$\degr$ to $\sim$37$\degr$, indicating a tendency of more radial propagation.
The value of $\phi_1$ almost keeps constant (-6$\degr$) during the propagation.
In the bottom panel of Figure~\ref{fig3}, 
the heliocentric distance ($d$) of the prominence leading front is drawn with blue pluses and red diamonds in the FOV of AIA and LASCO-C2, respectively.
A curve fitting using a quadratic function results in an acceleration of $\sim$978 m s$^{-2}$ during the initial eight minutes, 
which is cotemporary with the impulsive phase of the M3.5 flare as displayed in the top panel of Figure~\ref{fig3}.
A linear fitting during 08:00$-$08:48 UT in the decay phase of the flare results in a constant speed of $\sim$407 km s$^{-1}$, 
which is approximately one third of the apparent speed ($\sim$1186 km s$^{-1}$) of CME leading edge.

	\begin{deluxetable*}{rccccccc}
		\digitalasset
		\tablewidth{\textwidth}
		\tablecaption{Parameters of the cones during 07:25$-$08:48 UT,
		including the length ($r$) and angular width ($\omega$), inclination angles ($\theta_1$ and $\phi_1$), 
		the total height ($l$) and heliocentric distances ($d$) of leading fronts.
		\label{tab2}}
		\tablecolumns{8}
		\tablenum{2}
		\tablehead{
			\colhead{Time} &
			\colhead{Instrument} &
			\colhead{$r$} &
			\colhead{$\omega$} &
			\colhead{$\theta_1$} &
			\colhead{$\phi_1$} &
			\colhead{$l$} &
			\colhead{$d$}\\
			\colhead{(UT)} &
			\colhead{  } &
			\colhead{($\arcsec$)} &
			\colhead{($\degr$)} &
			\colhead{($\degr$)} &
			\colhead{($\degr$)} &
			\colhead{($\arcsec$)} &
			\colhead{($R_{\sun}$)}
		}
		\startdata
		07:25:44 & SDO/AIA & 57$\pm$3 & 14.5$\pm$1 & 60$\pm$0.5 & -6 & 64.71 & 1.03  \\
		07:26:44 & SDO/AIA & 65$\pm$3 & 15$\pm$1 & 58$\pm$0.5 & -6  & 74.12 & 1.03 \\
		07:27:44 & SDO/AIA & 74$\pm$3 & 17$\pm$1 & 57$\pm$0.5 & -6  & 85.88 &1.04 \\
		07:28:44 & SDO/AIA & 93$\pm$3 & 17.5$\pm$1 & 56.5$\pm$0.5 & -6  & 108.41 &1.05  \\
		07:29:44 & SDO/AIA & 109$\pm$5 & 19.5$\pm$1 & 55$\pm$0.5 & -6  &129.33 & 1.06 \\
		07:30:44 & SDO/AIA & 133$\pm$5 & 20.5$\pm$1 & 55 $\pm$0.5 & -6  &159.21 & 1.07 \\
		07:31:44 & SDO/AIA & 166$\pm$5 & 20$\pm$1 & 54.5$\pm$0.5 & -6  &197.83 & 1.09 \\
		07:32:44 & SDO/AIA & 217$\pm$5 & 20$\pm$1 & 54$\pm$0.5 & -6  & 258.61 & 1.12 \\
		07:33:44 & SDO/AIA & 290$\pm$7 & 20$\pm$1 & 53$\pm$0.5 & -6  & 345.61 & 1.17 \\
		07:46:46 &STB/EUVI & 533$\pm$10 & 20$\pm$1.5 & 52$\pm$1 & -6  & 635.21 & 1.34 \\
		08:00:00 & LASCO-C2& 1700$\pm$30 &13.7$\pm$1.5 & 39.3$\pm$1 & -6  &1916.44 & 2.41 \\
		08:12:00 & LASCO-C2& 2270$\pm$35 &14$\pm$1.5 & 38.7$\pm$1 & -6  & 2565.77 & 2.92 \\
		08:24:00 & LASCO-C2& 2770$\pm$50 &15$\pm$1.5 & 38.3$\pm$1 & -6  & 3158.58 & 3.35 \\
		08:36:00 & LASCO-C2& 3180$\pm$65 &15.3$\pm$1.5 & 38$\pm$1 & -6  & 3635.68 & 3.71 \\
		08:48:00 & LASCO-C2& 3630$\pm$90 &15.3$\pm$2 & 37.2$\pm$1 & -6  & 4150.17 & 4.12 \\
		\enddata
	\end{deluxetable*}
	
	\begin{figure} 
		\includegraphics[width=0.65\textwidth,clip=]{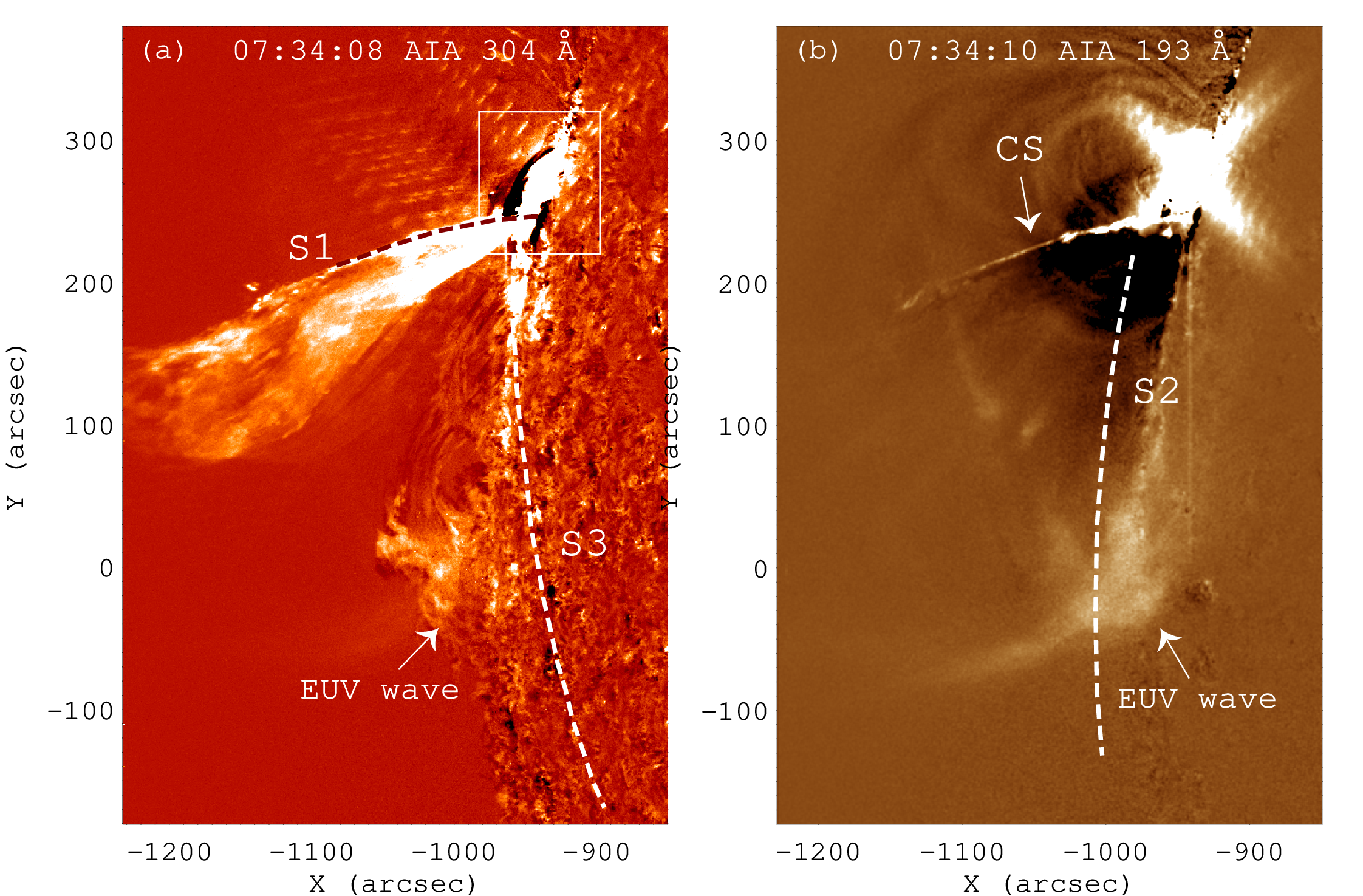}
		\centering
		\caption{The prominence and thin current sheet observed in AIA 304 and 193 {\AA} around 07:34:09 UT. 
		The curved slice S1 (brown dashed line) is used to investigate the evolutions of plasmoids along the current sheet.
		Two curved slices S2 and S3 (white dashed lines) are used to investigate the evolution of EUV wave.
		An animation showing the EUV wave observed by AIA 304 and 193 {\AA} is available.
		It covers a duration of 15 minutes from 07:24 UT to 07:39 UT on 2011 Feburary 24. The entire animation runs for $\sim$4 s.
		(An animation of this figure is available in the online article.)}
		\label{fig7}
	\end{figure}
	
	\begin{figure} 
		\includegraphics[width=0.83\textwidth,clip=]{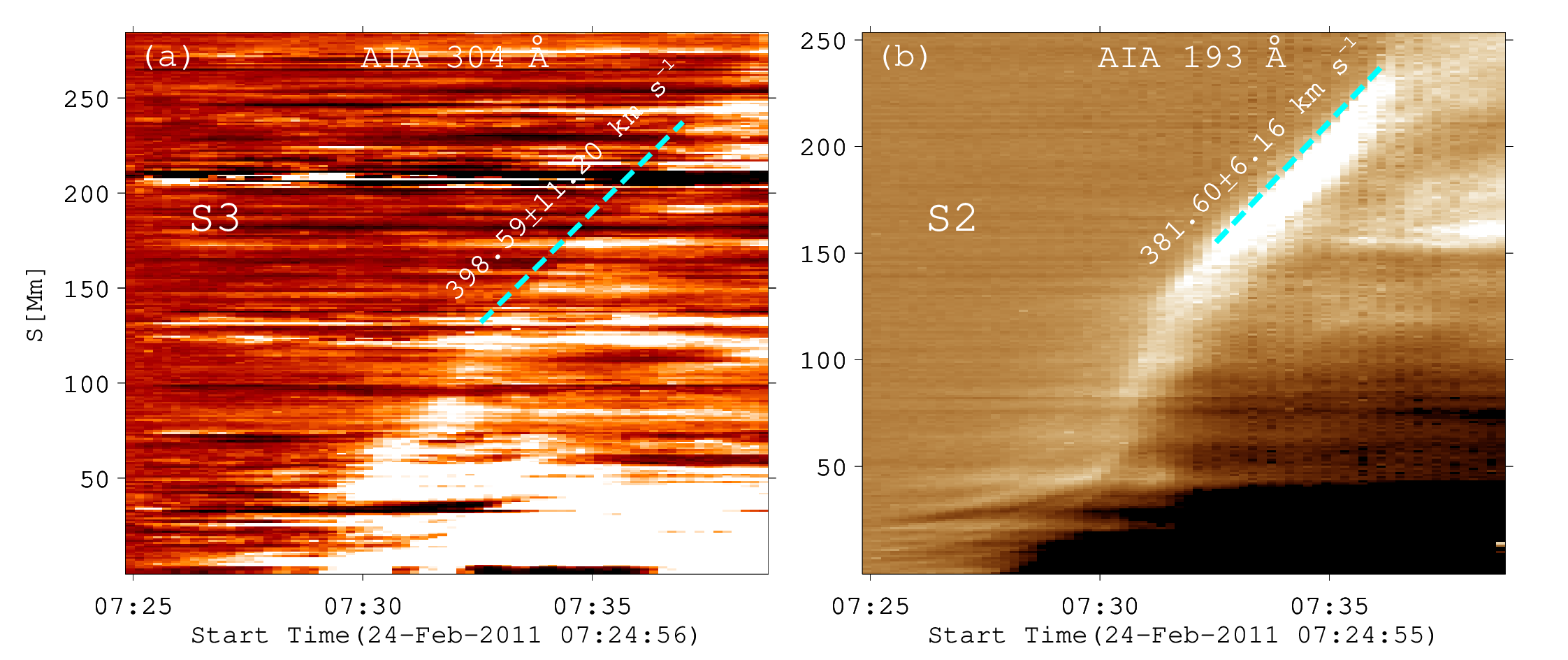}
		\centering
		\caption{Time-distance diagrams of S3 and S2 using the base-difference images in AIA 304 and 193 {\AA}.
		$s=254$ Mm and $s=285$ Mm stand for the south endpoints of S2 and S3.
		The speeds of EUV wave in two passbands are labeled.}
		\label{fig8}
	\end{figure}

The prominence eruption and associated CME generate an EUV wave propagating southward, rather than isotropically.
Figure~\ref{fig7} shows the AIA 304 and 193 {\AA} base-difference images around 07:34:09 UT (see also the online animation \textit{anim2.mp4}).
The white arrows point to the EUV wave. 
To calculate the speed of EUV wave, two curved slices (S2 and S3) with lengths of $\sim$254 Mm and $\sim$285 Mm 
along the direction of propagation are selected and drawn with white dashed lines.
Time-distance diagrams of S3 and S2 are displayed in the left and right panels of Figure~\ref{fig8}. 
The velocities of EUV wave in 304 and 193 {\AA} are calculated to be $\sim$398.59 and $\sim$381.60 km s$^{-1}$ during 07:30$-$07:36 UT.

\subsection{Plasmoids and Flare QPPs} \label{qpp}
\begin{figure}
\includegraphics[width=0.80\textwidth,clip=]{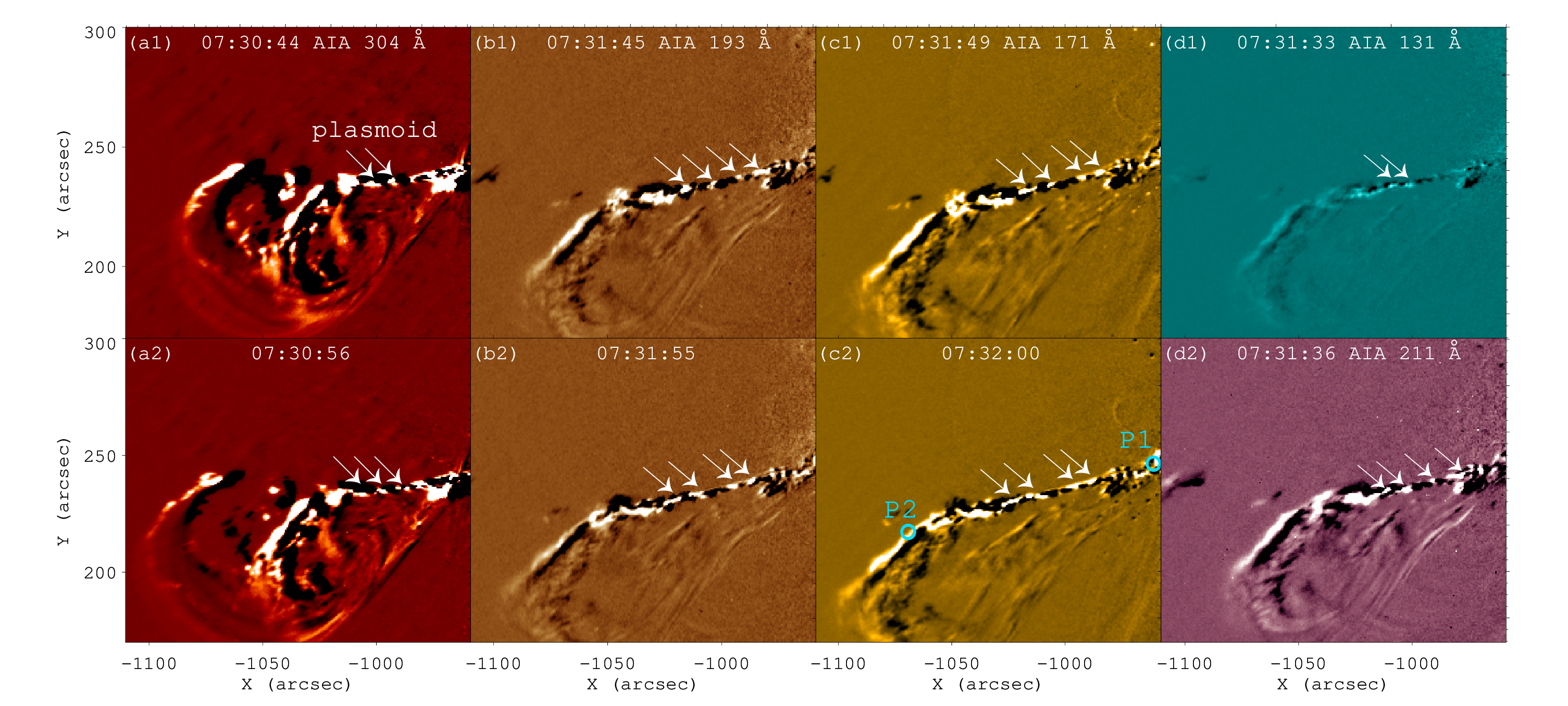}
\centering
\caption{Flare current sheet and plasmoids observed in AIA running-difference images, which are pointed by white arrows.
              In panel (c2), two blue circles (P1 and P2) mark endpoints of the current sheet.
              An animation showing the current sheet and plasmoids observed by AIA 304 and 193 {\AA} is available.
	     It covers a duration of 6 minutes from 07:29 UT to 07:35 UT on 2011 Feburary 24. The entire animation runs for $\sim$3 s.
	     (An animation of this figure is available in the online article.)}
\label{fig9}
\end{figure}

\begin{figure} 
		\includegraphics[width=0.70\textwidth,clip=]{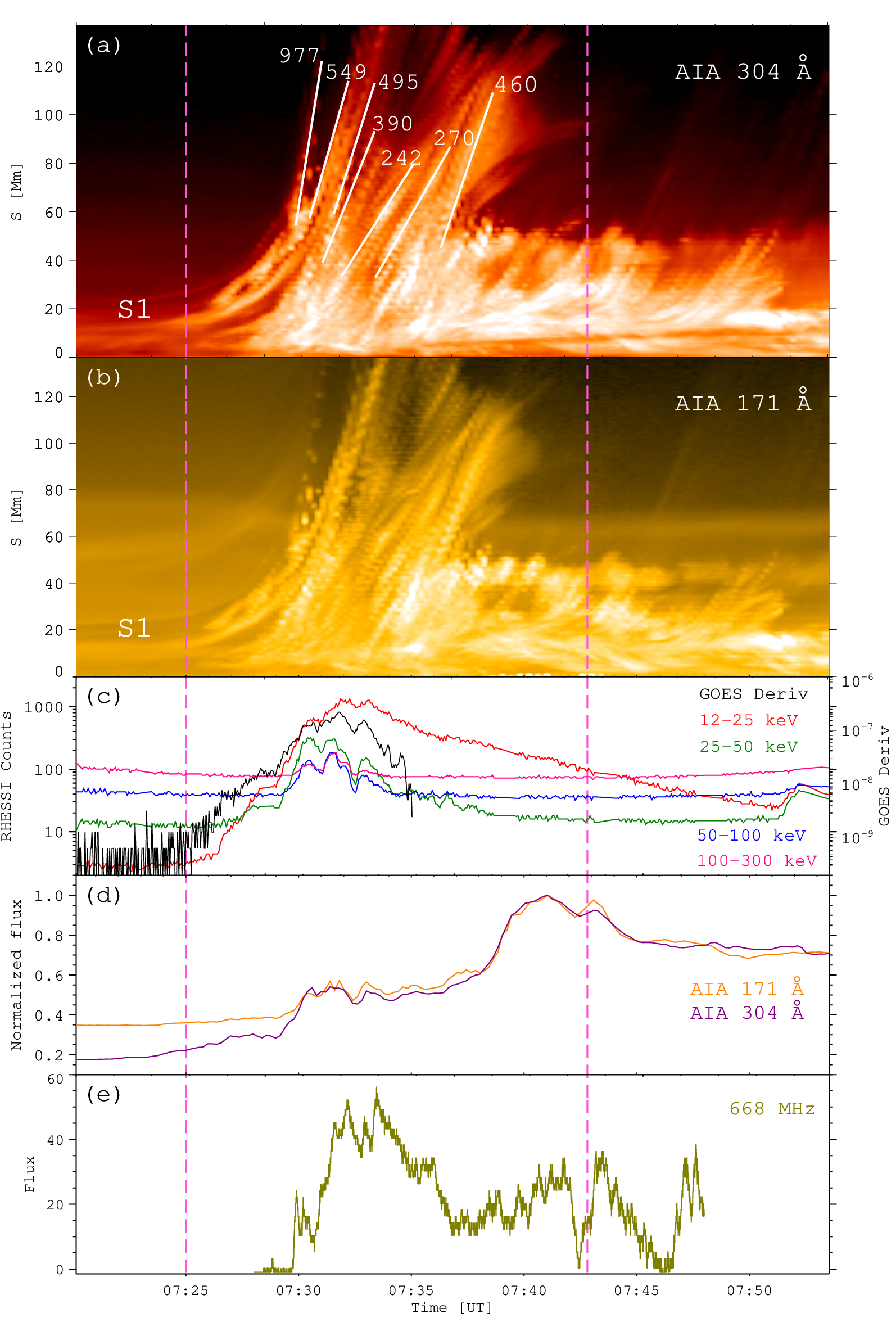}
		\centering
		\caption{(a)-(b) time-distance diagrams of slice S1 in 304 and 171 {\AA}.
		The velocities of rising plasmoids in units of km s$^{-1}$ are labeled.
		(c) GOES time derivative of 1$-$8 {\AA} fluxes (black line) and HXR fluxes of the flare at four energy bands.
		(d) Normalized fluxes of the flare in AIA 171 {\AA} (orange line) and 304 {\AA} (purple line).
		(e) Radio fluxes of the flare at 668 MHz.
		Two vertical dashed lines indicate the time interval during which the plasmoids appear.}
		\label{fig10}
	\end{figure}

As the prominence rises, a long and thin current sheet and plasmoids formed behind the CME.
To better show the faint, dynamic plasmoids within the current sheet, 
we utilize running-difference images, in which plasmoids are more evident than in the direct AIA images.
Figure~\ref{fig9} shows the snapshots of AIA running-difference images in 131, 171, 193, 211, and 304 {\AA} during 07:31$-$07:33 UT 
(see also the online animation \textit{anm3.mp4}).
The white arrows point to the bright and compact plasmoids, which move outward along the current sheet.
The appearance of plasmoids in multiwavelengths indicate their multithermal nature. 
In panel (c2), the blue circles mark the two endpoints (P1 and P2) of current sheet with heliocentric distances of 1.039 and 1.141 $R_{\sun}$, respectively.
To calculate the velocities of these plasmoids, we select a curved slice (S1) with a length of $\sim$137 Mm,
which is drawn with a brown dashed line in Figure~\ref{fig7}(a).
Time-distance diagrams of S1 in 304 and 171 {\AA} are displayed in Figure~\ref{fig10}(a)-(b).
Two vertical dashed lines delineate the start and end times of the plasmoids along the current sheet.
Around 07:25 UT, the first plasmoid appears, moves upward, and accelerates. The apparent velocity reaches up to $\sim$977 km s$^{-1}$.
Then, a series of plasmoids show up intermittently and move outward. 
The trajectories of another six plasmoids could be well identified with velocities between 240 and 550 km s$^{-1}$.

In Figure~\ref{fig10}(c), HXR light curves of the flare at four energy bands (12$-$25, 25$-$50, 50$-$100, and 100$-$300 keV) are plotted with red, green, blue, and pink lines.
Time derivative of the GOES flux in 1$-$8 {\AA} is plotted with a black line, 
which is well correlated with the 12$-$25 keV flux, an indication of the Neupert effect \citep{neu68,ver05}.
Moreover, the HXR pulsations are coincident with the formation and ejection of plasmoids during the impulsive phase of the flare.
\citet{ye23} performed full 3D MHD simulations of a CME and the large-scale current sheet between the flare and the flux rope.
The appearance of abundant, elongated plasmoids in the current sheet is cotemporary with the increase of reconnection rate and therefore the HXR emissions.
Meanwhile, the plasmoids experience splitting, merging, and kinking processes in a complex way.

\begin{figure} 
		\includegraphics[width=0.85\textwidth,clip=]{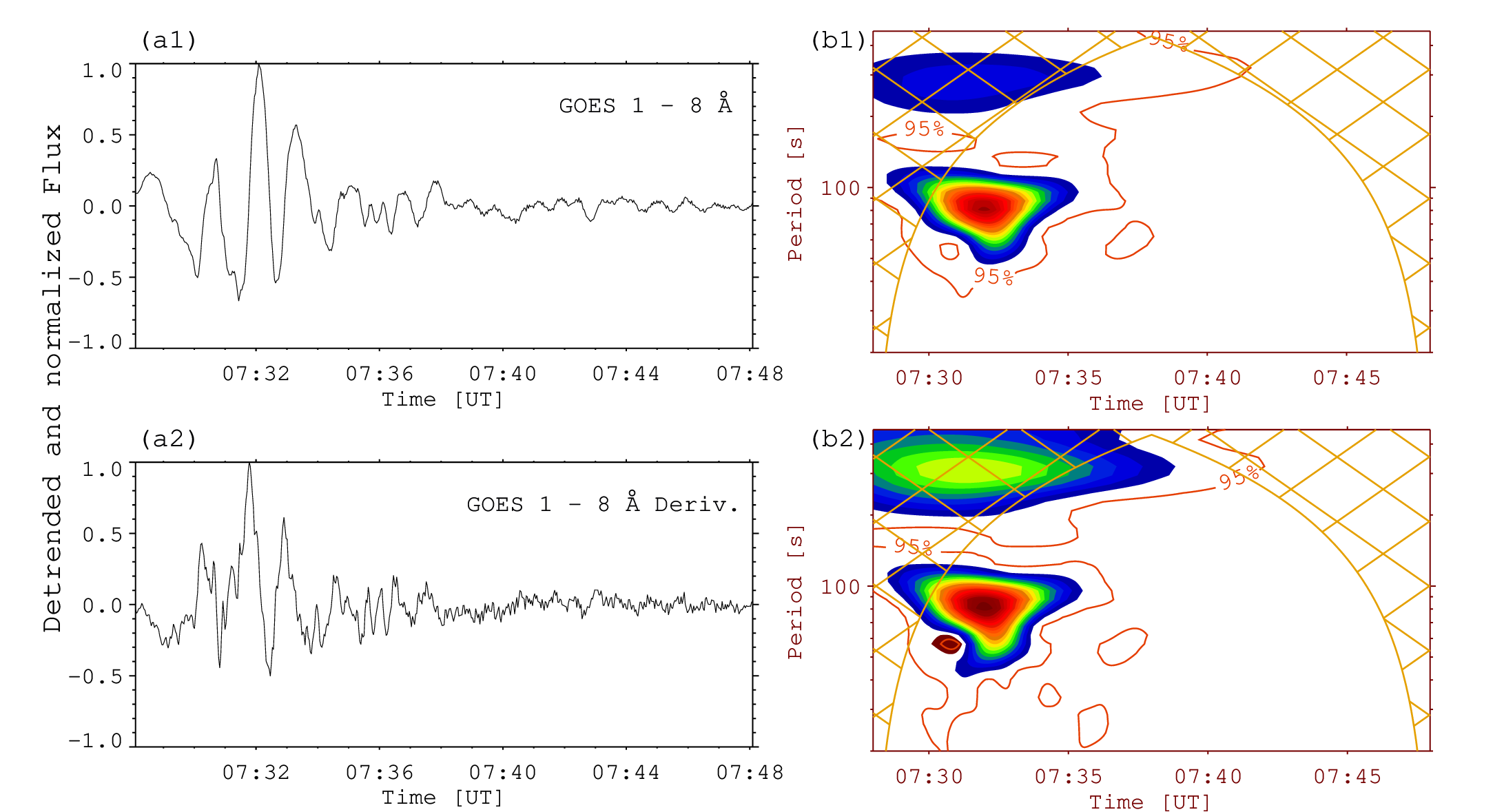}
		\centering
		\caption{Left panels: Detrended and normalized fluxes in GOES 1$-$8 {\AA} and time derivatives of 1$-$8 {\AA}.
		Right panels: Corresponding wavelet transforms of the fluxes.}
		\label{fig11}
	\end{figure}

\begin{figure} 
		\includegraphics[width=0.85\textwidth,clip=]{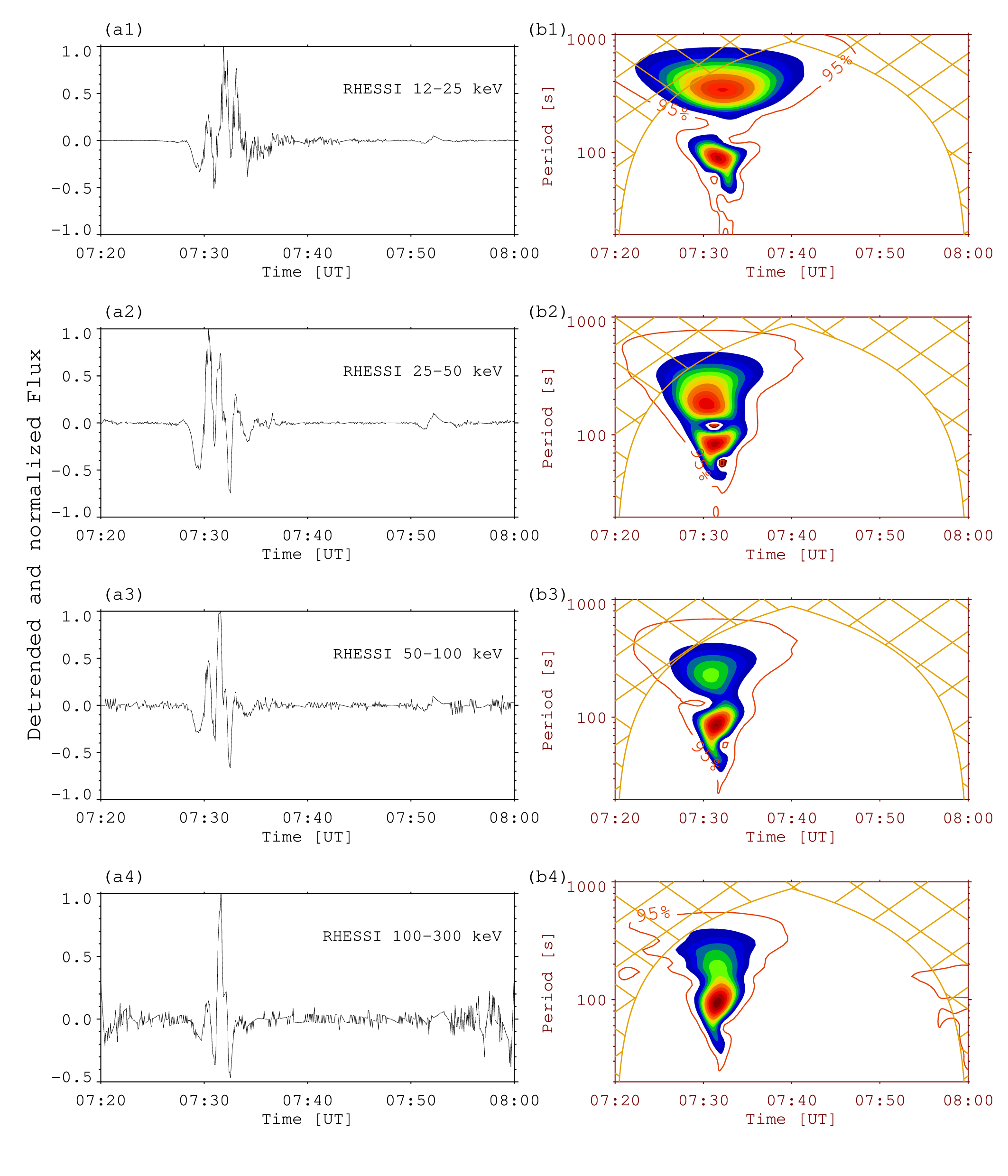}
		\centering
		\caption{Left panels: Detrended and normalized fluxes at four energy bands in HXR.
		Right panels: Corresponding wavelet transforms of the fluxes.}
		\label{fig12}
	\end{figure}

To investigate QPPs during the flare, the X-ray fluxes are detrended and normalized before carrying out the wavelet analysis.
In Figure~\ref{fig11}, the left panels show the normalized fluxes in 1$-$8 {\AA} and time derivative of the 1$-$8 {\AA}.
The wavelet transforms of these fluxes are displayed in the right panels. Both of them show clear QPP with periods of $\sim$80 s and $\sim$82 s, respectively.
Likewise, in Figure~\ref{fig12}, the left panels show detrended and normalized HXR fluxes at four energy bands (12$-$25, 25$-$50, 50$-$100, 100$-$300 keV).
The right panels show the corresponding wavelet transforms. The periods of QPPs in HXR lie in the range of 83$-$92 s (see Table~\ref{tab3}).
The period at 25$-$50 keV is the closest to that in SXR derivative.

\begin{deluxetable*}{ccc}
		\digitalasset
		\tablewidth{\textwidth}
		\tablecaption{Summary of QPPs of the M3.5 class flare detected in multiwavelengths.
			\label{tab3}}
		\tablecolumns{3}
		\tablenum{3}
		\tablehead{
			\colhead{Instrument} &
			\colhead{Waveband} &
			\colhead{Period} \\
			\colhead{} &
			\colhead{} &
			\colhead{(s)} 
		}
		\startdata
	         GOES & 1$-$8 {\AA}  & 80  \\
                 GOES & 1$-$8 {\AA} Deriv. & 82 \\
                 RHESSI & 12$-$25 keV & 89  \\
                 RHESSI & 25$-$50 keV & 83  \\
                 RHESSI & 50$-$100 keV & 87  \\
                 RHESSI & 100$-$300 keV & 92  \\
                 SDO/AIA & 171 {\AA}  &  94, 120 \\
                 SDO/AIA & 304 {\AA}  &  92  \\
                 BLEN7M & 668 MHz &  112$-$117 
		\enddata
	\end{deluxetable*}
	
\begin{figure} 
		\includegraphics[width=0.75\textwidth,clip=]{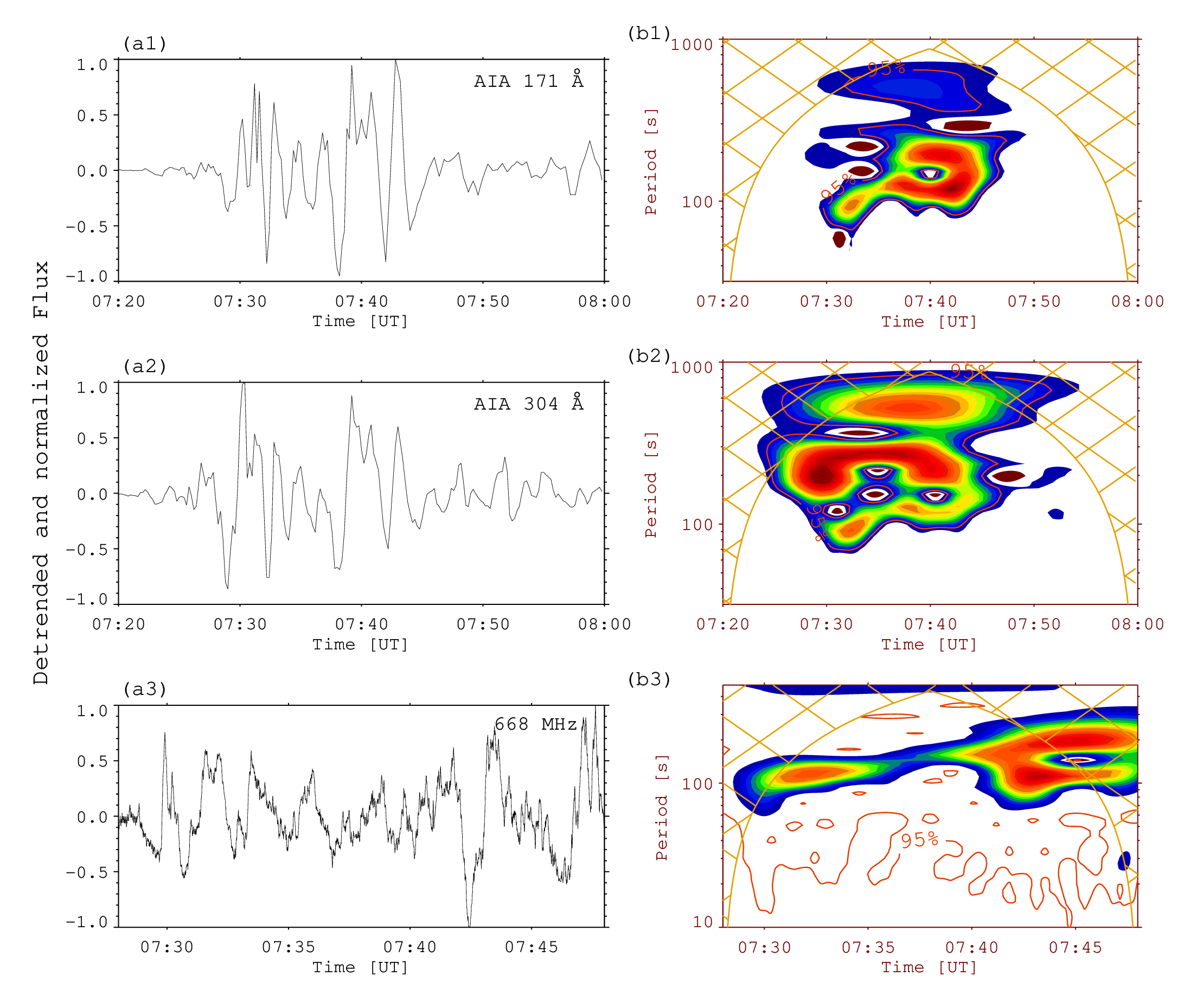}
		\centering
		\caption{Left panels: Detrended and normalized fluxes of AIA 171 Å, 304 Å, and e-Callisto/BLEN7M 668 MHz. 
		Right panels: Corresponding wavelet transforms of the fluxes.}
		\label{fig13}
	\end{figure}

\citet{lid15} studied the X1.6 class flare on 2014 September 10, which exhibits a 4-minute QPP in multiwavelengths.
To explore whether there are QPPs in EUV wavelengths during the M3.5 class flare, we derive the integrated intensities of the flare region, 
which is denoted with a white rectangle (85$\arcsec\times100\arcsec$) in Figure~\ref{fig7}(a).
The light curves of the flare in AIA 171 and 304 {\AA} are drawn with orange and purple lines in Figure~\ref{fig10}(d), respectively.
The corresponding detrended light curves are displayed in Figure~\ref{fig13}(a1)-(a2).
The results of wavelet transform are displayed in Figure~\ref{fig13}(b1)-(b2).
QPPs are present with periods of $\sim$92 s in 304 {\AA} and $\sim$94, $\sim$120 s in 171 {\AA}.
The periods (92 s and 94 s) in EUV are close to that of HXR at 12$-$25 and 100$-$300 keV.
Meanwhile, the period of $\sim$120 s is equal to that of speed oscillation of blob ``A" as a part of the eruptive flux rope \citep{kum12}.
	
\begin{figure} 
	\includegraphics[width=0.50\textwidth,clip=]{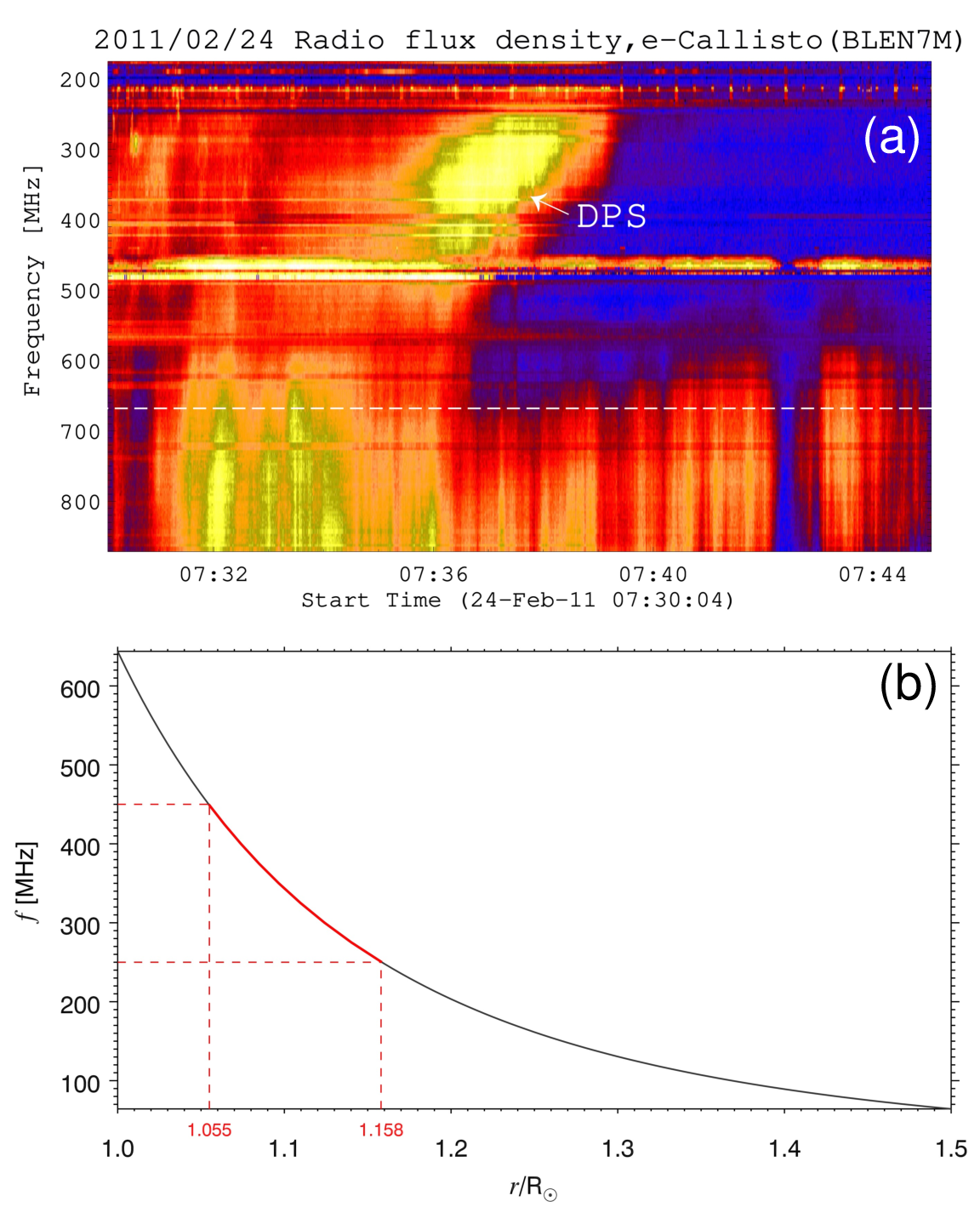}
	\centering
	\caption{(a) Radio dynamic spectra of the flare observed by e-Callisto/BLEN7M. The DPS is pointed by a white arrow.
	The horizontal dashed line denotes the frequency at 668 MHz.
	(b) Radial dependence of the local electron plasma frequency between 1.0 and 1.5 $R_{\sun}$.
	The red solid line denotes the region where the plasma frequency is between 250 and 450 MHz.}
	\label{fig14}
    \end{figure}

As shown in their Fig. 5 and Fig. 7 \citep{kum12}, the flare was associated with a type III radio burst, which was generated by nonthermal electrons propagating 
outward along open magnetic field lines \citep{zqm16,cla21}.
Figure~\ref{fig14}(a) shows the radio dynamic spectra recorded by e-Callisto/BLEN7M in the frequency range of 175$-$870 MHz, 
revealing the enhancement of radio emissions at decimeter wavelengths during the flare impulsive phase.
The radio fluxes at 668 MHz are extracted and plotted with an olive line in Figure~\ref{fig10}(e).
The detrended fluxes are plotted in Figure~\ref{fig13}(a3) and the wavelet transform is drawn in Figure~\ref{fig13}(b3).
The radio fluxes also present QPPs with periods of 112$-$117 s, which are comparable to that in 171 {\AA}. 
QPPs of the flare detected in multiwavelengths are summarized in Table~\ref{tab3}.
	
In Figure~\ref{fig14}(a), the white arrow points to the DPS whose frequency drifts from $\sim$450 to $\sim$250 MHz during 07:35$-$07:39 UT.
Considering that the emission mechanism of DPS is plasma emission in most cases \citep{dulk85}, 
the frequency drift indicates that the radio source of DPS moves outward from the Sun,
which is in agreement with the outward movements of plasmoids displayed in Figure~\ref{fig10}(a)-(b).
In Figure~\ref{fig14}(b), the radial dependence of the local electron plasma frequency ($f$) between 1.0 and 1.5 $R_{\sun}$ is drawn with a grey line.
Here, we adopt the electron density model \citep{mann99}:
\begin{equation}
 n_{e}(r)=N_{s}\times e^{c_{2}\times(1.0/r-1.0)},
\end{equation}
where $N_{s}=5.14\times10^9$ cm$^{-3}$, $c_{2}=13.83$, and $r$ denotes the heliocentric distance in units of $R_{\sun}$.
The plasma frequency (in units of MHz) is:
\begin{equation}
  f=8.98\times10^{-3}\times \sqrt{n_e}.
\end{equation}
It is seen from Figure~\ref{fig14}(b) that the value of $f$ decreases from 643.8 MHz at 1.0 $R_{\sun}$ to 64.2 MHz at 1.5 $R_{\sun}$.
The red solid line denotes the region where $f$ decreases from 450 MHz at 1.055 $R_{\sun}$ to 250 MHz at 1.158 $R_{\sun}$.
Therefore, the formation heights of DPS are satisfactorily consistent with the heights of plasmoids in the flare current sheet (Figure~\ref{fig9}(c2)).

\section{Discussion} \label{dis}
\subsection{Cause of nonradial prominence eruption}
Nonradial solar eruptions are frequently observed \citep{chen24,zqm24,sah25}. 
\citet{zqm21} studied two CMEs driven by nonradial prominence eruptions with inclination angles $\theta_1$ reaching up to 70$\degr$ and 60$\degr$
on 2011 August 11 and 2012 December 7.
\citet{zqm22a} studied the early evolution of a halo CME simultaneously observed by STEREO and SOHO on 2011 June 21, 
finding that the inclination angle $\theta_1$ decreases significantly from $\sim$51$\degr$ to $\sim$18$\degr$, while $\phi_1$ is almost invariant,
which is quite similar to the current study.
\citet{zqm24} analyzed the prominence eruption in the northwest direction on 2023 March 07. 
Interestingly, the absolute value of $\theta_1$ first increases from 36$\degr$ to 47$\degr$ before returning to 36$\degr$ and keeping up.
The cause of nonradial propagations of a prominence or a CME is the imbalance of magnetic energy density around it \citep{gui11}.
A prominence or a CME always inclines toward the region with lower magnetic energy density \citep{shen11}.
Specifically, open magnetic field lines at an adjacent coronal hole or at the boundary of ARs may strongly affect the propagation of a prominence \citep{zqm22a,zqm22c}.
In our case, open field is found to the north of AR 11163 using the potential field source surface modeling  \citep[see Fig. 7 in][]{kum12}.
Therefore, the prominence deflects southward and moves in the southeast direction (Figure~\ref{fig2}).
As the prominence rises up, the imbalance of energy density around the prominence may decrease with height.
Accordingly, the inclination angle ($\theta_1$) decreases as well.

\subsection{Cause of flare QPPs}
Plasmoids are extensively observed in solar flares and coronal jets \citep{zqm14,tang25}.
\citet{tak12} observed simultaneous inflow and outflow in a C4.5 flare on 2010 August 18. Multiple plasmoids are ejected bidirectionally along the current sheet.
The upward and downward velocities are 220$-$460 km s$^{-1}$ and 250$-$280 km s$^{-1}$, respectively.
\citet{xue16} detected a thin current sheet between a long eruptive filament and ambient chromospheric fibrils. 
The current sheet could be observed in AIA 304, 171, 335, and 131 {\AA}.
Besides, the differential emission measure analysis indicates that the current sheet is multithermal.
\citet{li16} noticed magnetic reconnection between an eruptive filament and its nearby coronal loops.
During the reconnection, plasmoids form and move bidirectionally along the current sheet. They are multithermal with average temperatures of a few MK.
In our study, the current sheet and plasmoids are clearly demonstrated not only in AIA EUV wavelengths (Figure~\ref{fig9}),
but also in AIA 1600 {\AA} \citep[][see their Fig. 3]{kum12}, which means that the current sheet and plasmoids are multithermal.
The flux rope holds up the prominence, which might take part in magnetic reconnection during its eruption.
Therefore, the current sheet and plamoids have both hot and cool components as observed in EUV and UV wavelengths.

QPPs are widespread in solar flares \citep{ing16,hay20,ing24}. \citet{hay20} analyzed more than 5500 flares, finding that $\sim$46\% of X-class, $\sim$29\% of M-class,
and $\sim$7\% of C-class flares present stationary QPPs. Besides, the period is positively correlated with the flare duration.
The physical origins of QPPs are still not fully understood. 
The most straightforward way of creating QPPs is quasi-periodic or intermittent magnetic reconnection and particle acceleration,
which could be spontaneous \citep{mur09,mc09} or modulated by MHD waves/oscillations \citep{naka06,naka11,zqm16}.
Another mechanism of QPPs is repetitive formation and coalescence of plasmoids in flare current sheets, which are tightly accompanied by DPS \citep{kli00,kar04}.  
\citet{lu22} reported observational features of tearing instability in the current sheet of an M7.7 flare on 2012 July 19.
Two plasmoids are clearly recognized and tracked along the current sheet, whose rising velocities are 640 and 1190 km s$^{-1}$.
Besides, the occurrence of DPS, whose frequency drifts from $\sim$580 to $\sim$200 MHz,
is coincident with the HXR pulsations at 50$-$100 keV and plasmoid ejections in the current sheet.
\citet{hua16} studied the same event, finding that QPPs with multiple periods exist in the HXR and radio emissions.
Specifically, the QPP in HXR emission (20$-$50 keV) has a single period of $\sim$270 s, 
while QPPs in radio emissions (17 GHz, 0.7$-$3 GHz) have double periods ($\sim$100 s and $\sim$270 s).
\citet{kou22} reported microwave QPPs with periods of 10$-$20 s and 30$-$60 s during the C5.9 class flare on 2017 July 13. 
Numerical simulations reveal that high-energy electrons are accelerated by quasi-periodic reconnections modulated by plasmoids within the current sheet.
\citet{kum25} presented direct imaging of the formation and ejection of multiple bidirectional plasmoids in the current sheets 
and the associated QPPs of the successive flares on 2015 April 22. QPPs with periods of $\sim$10 s and $\sim$100 s are detected in X-ray and radio wavelengths.
\citet{shi25} found two periods in the flare QPPs on 2022 August 26, a longer period of $\sim$2 minutes and a shorter period of $\sim$1 minute.
Meanwhile, repeated downward plasmoids are observed from the flare current sheet. 
Intermittent collisions and accelerations of electrons in these plasmoids 
could nicely explain QPPs in the flare looptop X-ray source and intensity fluctuations in the flare ribbons.
\citet{ye20} explored the role of turbulence for heating plasmas above the loop-top in eruptive flares.
Synthetic images and light curves of AIA EUV passbands show quasi-periodic radiation enhancement by turbulence when plasmoids move down and collide with the loop-top.
The period of QPP in 131, 171 and 304 {\AA} is 79.74 s, which is in agreement with the period in SXR in our study.
It is noted that QPPs with other periods ($\sim$67, $\sim$113, and $\sim$134 s) exist in their simulations.

Using MHD numerical simulations, \citet{tak17} discovered that the leading edge height of a flare-related CME (flux rope) 
presents a quasi-periodic fluctuation during the eruption. Consequently, the rising velocity of the CME (flux rope) varies periodically as well.
Similarly, in the 2.5 MHD simulations of a flux rope eruption, which leads to a CME and a flare, 
the outflow region in the current sheet becomes turbulent and the reconnection rate grows as the flux rope ascends.
The flux rope velocity increases in an oscillatory way \citep{ye19}.
In our study, Figure~\ref{fig9} and Figure~\ref{fig10} show intermittent plasmoids in the flare current sheet beneath the eruptive prominence.
Using a certain electron density model \citep{mann99}, the heights of DPS are basically accordant with the heights of current sheet,
suggesting that the DPS is indeed created by the electrons accelerated by the plasmoids moving upward in the current sheet.
As is shown in numerical simulations \citep{ye20,kou22}, the quasi-periodic magnetic reconnection and particle acceleration 
as a result of plasmoids in the thin current sheet could have multiple periods, rather than a single period. 
The periods from 80 to 120 s listed in Table~\ref{tab3} could result from the same process of tearing and plasmoid instability.
Quasi-periodic magnetic reconnection and particle acceleration with periods of 80$-$94 s would result in QPPs of the M3.5 flare detected in SXR, HXR, and EUV wavelengths.
Intermittent formations and propagation of plasmoids with a period of $\sim$120 s would result in quasi-periodic collisions 
and coalescence with the bottom of flux rope. Hence, the velocity of the flux rope oscillates with the same period \citep{kum12}.
Of course, the modulation of magnetic reconnection by the kink-mode oscillation of the rising flux rope with a period of $\sim$2 minutes could not be entirely ruled out.
Another important issue is the relationships between the QPP signals in different wavelengths, which may provide crucial information about the timing of particle acceleration 
and plasma heating. We will address this issue in the future investigation.
MHD simulations are urgently required to reproduce and explain the whole results of observations, which will be the major topic of our next paper.

\section{Summary} \label{sum}
In this paper, we report multiwavelength and multipoint observations of the prominence eruption originating from AR 11163, 
which is associated with an M3.5 class flare as well as a CME on 2011 February 24. The main results are summarized as follows:

\begin{enumerate}
   \item The rising prominence propagates nonradially in the southeast direction. 
   Using the revised cone model, 3D reconstructions of the icecream-like prominence reveal that
   the latitudinal inclination angle decreases from $\sim$60$\degr$ at 07:25 UT to $\sim$37$\degr$ at 08:48 UT, 
   suggesting that the prominence tends to propagate more radially. The longitudinal inclination angle almost keeps constant (-6$\degr$).
   The angular width of the prominence is between 13$\degr$ and 21$\degr$.
   The highly inclined prominence eruption and the related CME drive an EUV wave moving southward 
   at speeds of $\sim$381.60 km s$^{-1}$ and $\sim$398.59 km s$^{-1}$ observed in AIA 193 and 304 {\AA}.
   \item The flare shows QPPs in SXR, HXR, EUV, and radio wavelengths with periods of 80$-$120 s.
   A thin current sheet and multiple plasmoids are generated beneath the eruptive prominence, which are basically cotemporary with the flare QPPs.
   Combining with the appearance of DPS,
   the QPPs are most probably created by quasi-periodic magnetic reconnection and particle accelerations as a result of plasmoids in the flare current sheet.
   Our results may have implication on QPPs detected in stellar flares.
\end{enumerate}

\begin{acknowledgments}
The authors appreciate the reviewer for his/her valuable suggestions to improve the quality of this article.
We would like to thank Profs. Ivan Zimovets, Pengfei Chen, Jing Huang, Xiaoli Yan, Lei Ni, Zhike Xue, Jing Ye, and Leping Li for their helpful discussions.
SDO is a mission of NASA\rq{}s Living With a Star Program. AIA data are courtesy of the NASA/SDO science teams.
STEREO/SECCHI data are provided by a consortium of US, UK, Germany, Belgium, and France.
The e-Callisto data are courtesy of the Institute for Data Science FHNW Brugg/Windisch, Switzerland.
This work is supported by the National Key R\&D Program of China 2021YFA1600500 (2021YFA1600502), 2022YFF0503003 (2022YFF0503000),
the Strategic Priority Research Program of the Chinese Academy of Sciences, grant No. XDB0560000,
NSFC under the grant numbers 12373065, Natural Science Foundation of Jiangsu Province (BK20231510),
and Project Supported by the Specialized Research Fund for State Key Laboratory of Solar Activity and Space Weather.
\end{acknowledgments}

\bibliography{ref}
\bibliographystyle{aasjournalv7}

\end{document}